\documentclass{article}
\usepackage{spconf,amsmath,graphicx}
\usepackage{bm} 
\usepackage{booktabs,multirow,}
\usepackage{threeparttable, pifont}
\title{Multi-objective Progressive Clustering for Semi-supervised Domain Adaptation in Speaker Verification}

\name{Ze Li$^{1,2}$,Yuke Lin$^{1,2}$, Ning Jiang $^{3}$, Xiaoyi Qin$^{1,2}$,  Guoqing Zhao$^{3}$, Haiying Wu$^{3}$, Ming Li$^{1,2}$ \thanks{Corresponding Author: Ming Li.}}
\address{
$^1$School of Computer Science, Wuhan University, Wuhan, China \\
$^2$Suzhou Municipal Key Laboratory of  Multimodal Intelligent Systems,\\ Duke Kunshan University, Kunshan, China \\
$^3$Mashang Consumer Finance Co., Ltd, China \\
ming.li369@dukekunshan.edu.cn
}

\begin{document}
\ninept
\maketitle
\begin{abstract}
Utilizing the pseudo-labeling algorithm with large-scale unlabeled data becomes crucial for semi-supervised domain adaptation in speaker verification tasks. In this paper, we propose a novel pseudo-labeling method named Multi-objective Progressive Clustering (MoPC), specifically designed for semi-supervised domain adaptation. Firstly, we utilize limited labeled data from the target domain to derive domain-specific descriptors based on multiple distinct objectives, namely within-graph denoising, intra-class denoising and inter-class denoising. Then, the Infomap algorithm is adopted for embedding clustering, and the descriptors are leveraged to further refine the target domain's pseudo-labels. Moreover, to further improve the quality of pseudo labels, we introduce the subcenter-purification and progressive-merging strategy for label denoising. Our proposed MoPC method achieves 4.95\% EER and ranked the 1$^{st}$ place on the evaluation set of VoxSRC 2023 track 3. We also conduct additional experiments on the FFSVC dataset and yield promising results.

\end{abstract}

\section{Introduction}

\begin{figure*}
  \centering
  \includegraphics[width=\linewidth]{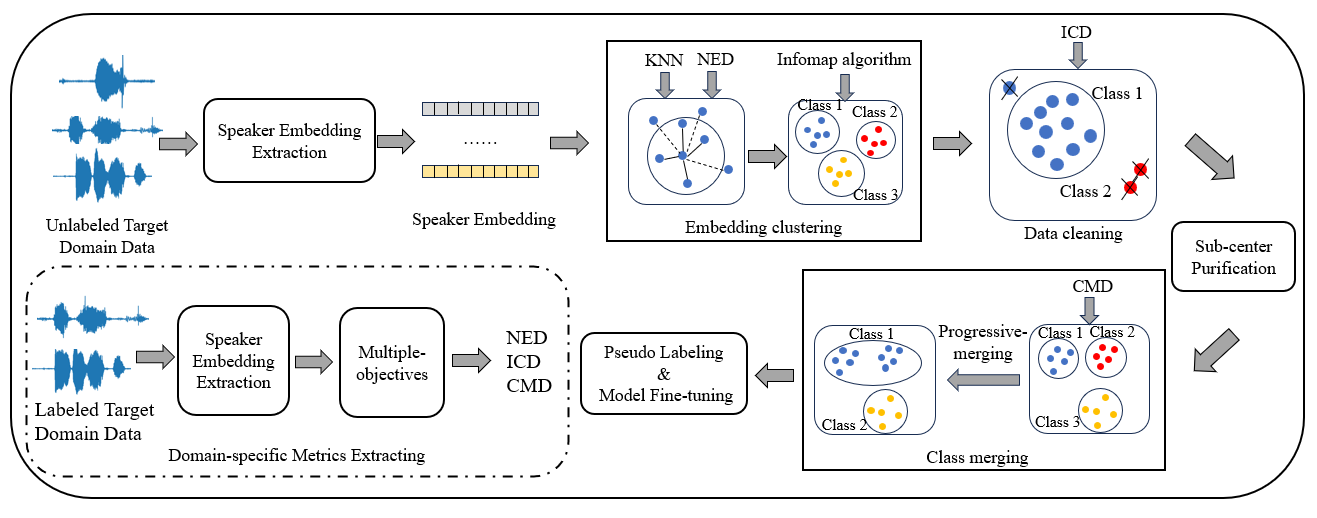}
  \caption{System framework of our MoPC method.}
  \label{fig:system_framework}
\end{figure*}

% ------------------Para.1-----------------------
Automatic speaker verification (ASV) aims to verify the identities of speakers from their voice samples. In recent years, thanks to the advancement of deep neural networks (DNN), deep learning-based SV systems, such as \cite{x-vector, ecapa}, have demonstrated impressive performance under different scenarios. Nevertheless, when an ASV system well trained on the source domain is applied to another domain, the performance may degrade significantly. Therefore, domain adaptation is an important and challenging issue in speaker verification.

To address this problem, researchers propose various methods to bridge the domain gap. These methods can be broadly categorized into statistical alignment \cite{coral, coral+},  generative adversarial \cite{gan}, self-supervised learning \cite{self-supervised} and pseudo-labeling \cite{dku-voxsrc22, zkd-voxsrc22}. Statistical alignment and generative adversarial approaches mainly focus on leveraging domain distribution information to bridge the domain gap. However, this often leads to overlooking finer-grained details and class-specific information. Self-supervised learning based methods utilize the correlation within target domain data, but the system performance still needs to be improved. Pseudo-labeling based method achieve competitive results in ASV domain adaptation tasks. It iteratively utilizes clustering algorithms to assign pseudo-labels to unlabeled target domain data and subsequently uses these pseudo-labels for supervised learning. However, the accuracy of the pseudo-labels largely depends on the robustness of the clustering algorithm and the reliability of the source domain's pre-trained model.

Many existing works focus on unsupervised domain adaptation, assuming the absence of labels in the target domain. However, considering that manually annotating a small amount of target data is easy to achieve, semi-supervised domain adaptation emerges as a more realistic challenge, which aims to learn target models from a few labeled target domain data and a large amount of unlabeled target domain data. For pseudo-labeling approaches, a straightforward method is to merge limited labeled data with large amount of pseudo-labeled data together for training. However, we believe that the labeled target domain data contains valuable domain-specific information, which aids in filtering out noisy labels from the pseudo-labeled data, thereby enhancing the performance of domain adaptation.

% 感觉intro不用讲那么多，说的有些详细了
In this study, we propose a novel pseudo-labeling method named Multi-objective Progressive Clustering (MoPC) for semi-supervised domain adaptation. To begin with, we define multiple objectives, including within-graph denoising, intra-class denoising and inter-class denoising. These objectives can be quantified from a small amount of labeled data to improve the embedding clustering quality of unlabeled data. Furthermore, we also introduce the subcenter-purification strategy to remove noisy classes and adopt the progressive-merging strategy to further enhance the quality of pseudo-labels.

\section{Methods}
\subsection{Domain-specific Descriptor Extraction}
\label{sec:multiple-objectives}
To enhance the quality of graph clustering, we define multiple objectives: within-graph denoising, intra-class denoising and inter-class denoising. Then, we utilized the limited labeled data from the target domain to derive the domain-specific descriptors (denoted as Noise-Edge Descripter($NED$), Intra-Class Descripter($ICD$) and Class-Merging Descripter($CMD$), respectively) based on these objectives. The methods of deriving these descriptors are described as follows:

\textbf{Descriptor} $\bm{NED}$ \textbf{.} We compute the pairwise cosine similarity between each embedding and those from different classes, then select the maximum value as $NED$:

\begin{equation}
    NED = \max_{\substack{1 \leq i \neq j \leq K \\ 1 \leq a \leq N_{i} , 1 \leq b \leq N_{j}}  }cosine\left(\mathbf{z}_{i, a}, \mathbf{z}_{j,b}\right)
\end{equation}
where $\mathbf{z}_{i,a}$ is the $a^{th}$ embedding of the $i^{th}$ class, $K$ represents the total number of classes and $N_{i}$ denotes the total number samples of the $i^{th}$ class. The embeddings are extracted by the model trained on the source domain. 

\textbf{Descriptor} $\bm{ICD}$ \textbf{.} We calculate the cosine similarity between the embedding of each class and its respective centroid vector and select the maximum one from the minimum values of each class as Descriptor $ICD$:

\begin{equation}
    ICD = \max_{\substack{1 \leq i \leq k}} \min_{\substack{1 \leq a \leq N_{i}}} cosine\left(\mathbf{z}_{i, a}, \mathbf{C}_{i}\right)
\end{equation}
where $\mathbf{C}_{i}$ is the centroid vector of the $i^{th}$ class.

\textbf{Descriptor} $\bm{CMD}$ \textbf{.} We compute the cosine similarity between the centroid vectors of each class, then select the maximum one as descriptor $CMD$:

\begin{equation}
    CMD = \max_{\substack{1 \leq i \neq j \leq k}} cosine\left(\mathbf{C}_{i}, \mathbf{C}_{j}\right)
\end{equation}

\subsection{Overall Description of the MoPC Framework}
This section describes the proposed overall framework of our MoPC pseudo-labeling method for semi-supervised domain adaptation. The framework is shown in Fig \ref{fig:system_framework} and the steps are described as follows:

\noindent \textbf{Step 1: Speaker embedding extraction.} 
    Firstly, we utilize the model trained on the source domain as a feature extractor, to obtain speaker embeddings from target domain data. To maintain stability, utterances less than one second are discarded.
    
\noindent \textbf{Step 2: Embedding clustering.} 
    We generate the graph with speaker embedding using the $K$-Nearest Neighbors ($K$NN) algorithm, the parameter $K$ is determined by 'elbow' method \cite{cdw}. 
    Descriptor NED represents the highest similarity among embedding that do not belong to the same class. Therefore, it is highly probable that embeddings with similarity higher than NED belong to the same class. Hence, we retain edge weights greater than NED to eliminate within-graph noise. Then, the Infomap \cite{infomap} algorithm is employed for clustering based on the graph.

\noindent \textbf{Step 3: Data cleaning.} 
    Descriptor ICD represents the highest similarity of edge embeddings to their respective class centers. Therefore, embeddings with similarity higher than ICD to their class center are highly likely to belong to that class. Hence, we retain embeddings with similarity higher than ICD to their class center to eliminate intra-class noise. % 这句是不是有点啰嗦？
    Additionally, eliminate classes with deficient numbers of instances. % 这个原因怎么解释？数据量很少的类极大概率是噪声点？有什么理论支撑？
    
\noindent \textbf{Step 4: Sub-center purification.} 
    To further improve the quality of pseudo labels, we introduce a subcenter-purification strategy, to remove noisy classes by the soft labels output from Sub-Center head. For more detailed information, please refer to Section \ref{sec:subcenter-purification}.
    
\noindent \textbf{Step 5: Progressive class merging.} 
    To reduce inter-class noise, we employ a progressive-merging strategy to merge classes with the same ground-truth label while making every effort to avoid the increase in intra-class noise caused by erroneous merges. For more detailed information, please refer to Section \ref{sec:progressive-merging}.

\noindent \textbf{Step 6: Pseudo labeling and model fine-tuning.} 
    After completing the aforementioned steps, we assign pseudo-labels to unlabeled target domain data based on the clustering results. Then, both the unlabeled data with pseudo-labels and the labeled data are used for fine-tuning the pre-trained speaker model.

\subsection{Sub-center Purification}
\label{sec:subcenter-purification}
\begin{figure}
  \centering
  \includegraphics[width=\linewidth]{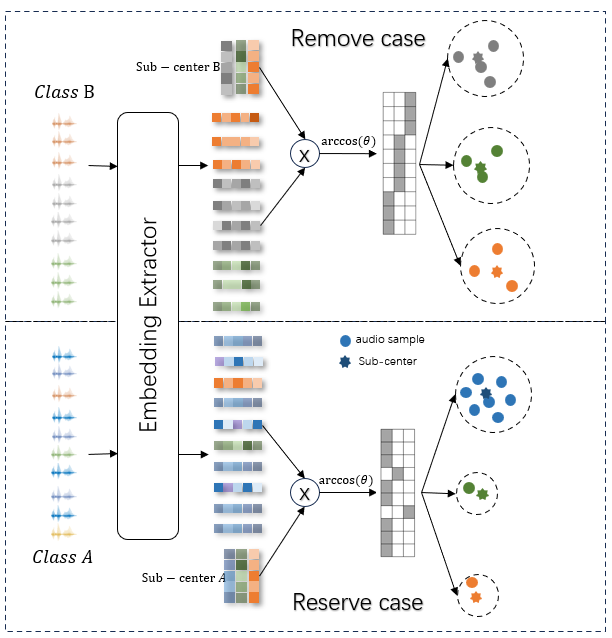}
  \caption{Sub-center Purification. Noisy data in one class tends to be uniformly distributed into various sub-centers (Top), while purer class data is more likely to select a distinct sub-center (Bottom).}
  \label{fig:subcenter}
\end{figure}

To address the issue of class impurity, inspired by the sub-center ArcFace \cite{subcenter-arcface} loss, we introduce the subcenter-purification strategy. In this process, we begin by assigning pseudo-labels to the unlabeled data after Step 3, using these as the input to train a Sub-center ArcFace classifier. After convergence, we stop training and pass all the data through the classifier to compute the selection probability for each class's sub-center. As shown in Fig \ref{fig:subcenter}, most of the highly pure class data will tend to choose one specific sub-center, while the class with lower purity will exhibit multiple sub-centers, and we could consider removing it.

\subsection{Progressive Class Merging}
\label{sec:progressive-merging}
\begin{figure}
  \centering
  \includegraphics[width=\linewidth]{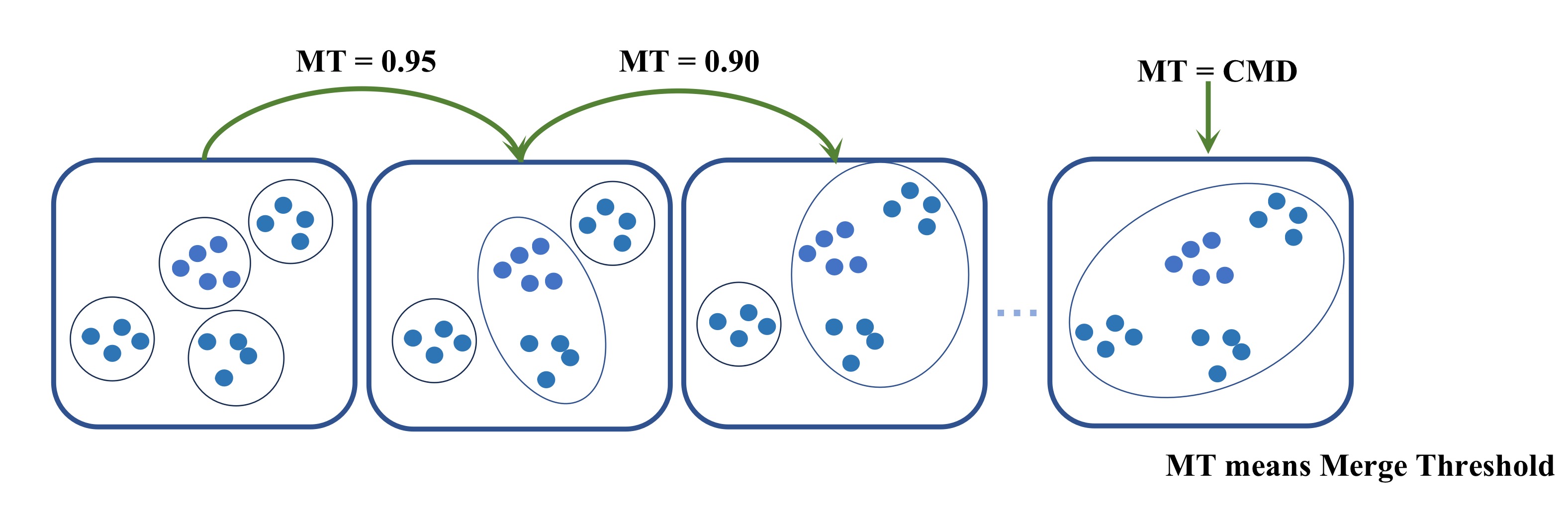}
  \caption{Progressive class merging. We set a series of descending threshold values to merge classes with high center similarities iteratively.}
  \label{fig:merge}
\end{figure}

During the clustering process, there may be cases where multiple classes share the same ground-truth label. In such instances, it is necessary to merge these classes; however, we also need to avoid the increase in intra-class noise caused by incorrect merges. To achieve this, inspired by the Agglomerative Hierarchical Clustering, we introduce the progerssive-merging strategy. As shown in Fig \ref{fig:merge}. Firstly, we set a series of merge thresholds ranging from higher value down to the descriptor CMD. Then, for each merge threshold, we merge those classes that their class center similarities satisfy the threshold condition and are each other's closet neighbors.

\begin{table*}[htbp]\centering 
    \scriptsize %\footnotesize \scriptsize
    \caption{The results of different backbone systems using our MoPC based semi-supervised learning framework on the  VoxSRC 22/23 Track 3 datasets.}
     \label{tab:cnceleb}
\begin{threeparttable}
    \begin{tabular}{lccccccc}
    \toprule
    \multirow{2}*{\textbf{ID \& Model}} & \multicolumn{2}{c}{\textbf{VoxSRC23 val}} & \multicolumn{2}{c}{\textbf{VoxSRC23 test}} & \multicolumn{2}{c}{\textbf{VoxSRC22 test}}\\
    \cmidrule(lr){2-3} \cmidrule(lr){4-5} \cmidrule(lr){6-7}& \textbf{EER[\%]} & \textbf{mDCF$_{0.05}$} & \textbf{EER[\%]} & \textbf{mDCF$_{0.05}$} & \textbf{EER[\%]} & \textbf{mDCF$_{0.05}$} \\
    \midrule 
    1 SimAM-ResNet100-ASP & 7.490 & 0.342 & 5.287 & 0.3037 & 6.927 & 0.409 \\
    2 ResNet100-TSP & 7.350 & 0.360 & - & - & - & - \\
    3 SimAM-ResNet100-ASP(v2) & 7.525 & 0.335 & - & - & - & - \\
    4 ResNet152-ASP & 7.240 & 0.347 & - & - & - & - \\
    5 ResNet152-TSP & 7.535 & 0.358 & - & - & - & - \\
    \midrule
    Fusion(1+2+3+4+5) & 6.725 & 0.311 & 4.952 & 0.2777 & 6.584 & 0.374 \\
    Fusion \cite{zkd-voxsrc22} * & - & - & - & - & 7.030 & 0.388 \\
    Fusion \cite{dku-voxsrc22} \ding{61} & - & - & - & - & 7.153 & 0.389 \\
    \bottomrule
    \end{tabular}
    \begin{tablenotes}
        \item * PGMVG clustering method, 1$^{st}$ place in VoxSRC 2022 track 3
        \item \ding{61} $k$-means clustering method, 2$^{st}$ place in VoxSRC 2022 track 3
    \end{tablenotes}
\end{threeparttable}
    
\end{table*}

\section{EXPERIMENTS}
\subsection{Dataset}
We employ two datasets from different languages or scenarios in our experiments to perform domain adaptation. 
The VoxCeleb2 \cite{vox2dev} dev set with speaker labels is used as the labeled data form the source domain. For unlabeled target domain data, we use the following two datasets: a subset of the CNCeleb2 \cite{cnceleb2} dev set provided by VoxSRC23 challenge \footnote{http://mm.kaist.ac.kr/datasets/voxceleb/voxsrc/competition2023.html}, the FFSVC2020 supplementary set provided by FFSVC2022 challenge \cite{ffsvc2022}. To perform semi-supervise domain adaptation, we further select the small subset of CNCeleb (1,000 utterances from 50 speakers) provided by VoxSRC23 challenge as the labeled target domain data for CNCeleb2, and randomly select 1,000 utterances from 10 speakers in the FFSVC2020 supplementary set as the labeled target domain data for FFSVC.

For evaluation, we use the offcial validation and test trial list of the VoxSRC23 Track 3 task and the official development trial list of FFSVC2022 to evaluate the systems trained in the semi-supervised manner.

\subsection{Model Usage}
We utilize two ResNet-Based systems for model training, including the ResNet100-based one \cite{idrd_voxsrc22} and the ResNet152-based one \cite{li2023dku}. In addition, we also adopte the SimAM \cite{simam} modules in backbone block. The acoustic features are 80-dimensional log Mel-filterbank energies with a frame length of 25ms and a hop size of 10ms. The extracted features are mean-normalized before feeding into the deep speaker network. We also introduce the temporal statistic pooling (TSP) and the attentive statistic pooling (ASP) \cite{asp_pooling} to obtain fix-dimension vectors. 

\subsection{Training Details}

Our training process can be divided into two stages, including (i) Supervised learning on the source domain (ii) Semi-supervised domain adaptation learning on the target domain.

\subsubsection{Supervised Learning Settings}
Since the source domain data is labeled, it can be used to train a feature extractor. In this phase, we adopt the on-the-fly data augmentation \cite{on-the-fly} to add additive background noise or convolutional reverberation noise for the time-domain waveform. The speed perturbation \cite{dku_voxsrc20}, which speeds up or down each utterance by a factor of 0.9 or 1.1, is applied to yield shifted pitch utterances that are considered from new speakers. The SGD optimizer with a momentum of 0.95 and weight decay of 1e-4 is used. We adopt the StepLR scheduler with 15 epochs decay. The init learning rate starts from 0.1, the minimum learning rate is 1.0e-4, and the decay factor is 0.1. The margin and scale of ArcFace are set as 0.2 and 32, respectively. We perform a linear warm-up learning rate schedule at the first 5 epochs to prevent model vibration and speed model training. The input frame length is fixed at 200 frames.

\subsubsection{Semi-Supervised Learning Settings}
In this stage, labeled and pseudo-labeled target domain data are used to fine-tune the pre-trained model. Only speaker augmentation with speed perturbation is retained for data augmentation, and the ArcFace is replaced by the Sub-center ArcFace. The learning rate starts from 1.0e-3 and gradually drops till convergence. 

Considering the size of the target domain data, we employ distinct fine-tuning strategies for VoxSRC Track3 and FFSVC. For VoxSRC Track3, the source model is directly fine-tuned using the target domain data. In contrast, the FFSVC dataset, with its fewer speakers, demands a cautious approach to avoid overfitting. Hence, we set the initial learning rate to 1.0e-5 and apply the Mix-FT strategy\cite{farfield_xiaoyi}. Additionally, for experiments on the VoxSRC23 dataset, AS-norm and QMF mentioned in \cite{li2023dku} are used for score calibration.

\section{Results}
Table \ref{tab:pseudo_label} displays the effectiveness of each step of our MoPC method. We 
utilize the ground-truth labels of the target domain data and the following metrics to access the quality of pseudo-labels: intra-class noise rate, inter-class noise rate, number of pseudo-label classes, number of ground-truth classes, and NMI \cite{cdw}. As we can see, in the scenario where only the Infomap algorithm is applied as the primary clustering method, both the intra-class noise and inter-class noise rates are quite high, reaching 11\% and 34.4\%, respectively. After we introduce the descriptor NED to eliminate within-graph noise, we can observe a significant decrease in both intra-class noise rate and inter-class noise rate, and the NMI has shown some improvement. Furthermore, due to the fact that some classes have very few samples, after removing the within-graph noise, the embeddings of these classes cannot form edges with embeddings from other classes and are ultimately treated as isolated points and removed. Consequently, the number of ground-truth classes decreases by over one hundred. After introducing the descriptor ICD and subcenter-purification strategy to improve the quality of pseudo-labels further, we can see an improvement in all metrics. After introducing the descriptor CMD with the progressive-merging strategy to eliminate inter-class noise, we see a significant decrease in inter-class noise rate, which has dropped from 28.5\% to 17.1\%. Due to some inevitable erroneous merges, the metric intra-class noise rate has increased from 1.71\% to 7.7\%. However, at this point, the number of pseudo-label classes is closer to the number of ground-truth label classes, the improvement in the performance of the inter-class noise rate far outweighs that of the intra-class noise rate, and the NMI metric has not decreased significantly. Therefore, we believe that the quality of pseudo-labels at this point is superior to the previous state.

Table \ref{tab:cnceleb} reports the results on different VoxSRC 2022 and 2023 Track 3 evaluation sets. As we can see, compared to both PGMVG and $k$-means methods, our approach is significantly ahead. Even on the SimAM-ResNet100-ASP single-system, our model outperforms the results achieved by their fused model on the VoxSRC 2022 Track 3 evaluation set. Furthermore, our fused system achieved 4.952\% EER and ranked the 1$^{st}$ place on the VoxSRC 2023 Track 3 evaluation set. 

Additionally, we conducted experiments on the FFSVC dataset to assess the universality of our method.  It can be observed in Table \ref{tab:ffsvc} that our method is equally effective on the FFSVC dataset. Compared to the $k$-means method, our method exhibits an improvement of nearly 20\% on the FFSVC 2022 development set.

\begin{table}[htbp]\centering \scriptsize
    \caption{NR$_1$ and NR$_2$ indicates the Intra-class and Inter-class noise rate, respectively. $P$ and $GT$ are Pseudo label and ground truth label.  }
     \label{tab:pseudo_label}
    \begin{tabular}{lccccccccc}
    \toprule
    \textbf{Method} & \textbf{NR$_1$[\%]} &\textbf{NR$_2$[\%]} & \textbf{Spk. of P} & \textbf{Spk. of GT} & \textbf{NMI} \\
    \midrule 
     k-means & 12.3 & 42.6 & 1800 & 1807 & 0.9179 \\
     \midrule 
     Based (Infomap) & 11.0 & 34.4 & 2128 & 1807 & 0.9412 \\
     + NED & 2.4 & 29.9 & 2080 & 1642 & 0.9783 \\
     ++ ICD & 2.2 & 29.7 & 2079 & 1636 & 0.9794 \\
     +++ subcenter & 1.7 & 28.5 & 2070 & 1626 & 0.9827 \\
     ++++ CMD & 7.7 & 17.1 & 1705 & 1626 & 0.9811 \\
    \bottomrule
    \end{tabular}
\end{table}

\begin{table}[htbp]\centering \scriptsize
    \caption{The results of the SimAM-ResNet100-ASP single system on the FFSVC22 development set.}
     \label{tab:ffsvc}
    \begin{tabular}{lccccccccc}
    \toprule
	\multirow{2}*{\textbf{Method}} & \multicolumn{2}{c}{\textbf{FFSVC-dev}} \\
	\cmidrule(lr){2-3} \cmidrule(lr){4-5}& \textbf{EER[\%]} & \textbf{mDCF$_{0.01}$} \\
    \midrule 
     Based & 10.097 & 0.736\\
    + $k$-means & 5.147 & 0.523 \\
    + MoPC(ours) & 4.125 & 0.461  \\
    \bottomrule
    \end{tabular}
\end{table}

\section{Conclusion}
In this paper, we proposed a novel pseudo-labeling method named Multi-objective Progressive Clustering (MoPC) for semi-supervised domain adaptation. The differences between our semi-supervised approach and previous methods lies in the more extensive utilization of the limited amount of target domain labeled data. The subcenter-purification and progressive-merging strategy are also introduced to improve the quality of pseudo labels further. Experimental results on CNCeleb2, FFSVC show that our proposed method is effective on different datasets.

\section{Acknowledgement}
This research is funded in part by the National Natural Science Foundation of China (62171207), Science and Technology Program of Suzhou City(SYC2022051) and MaShang Consumer Finance Co.Ltd. Many thanks for the computational resource provided by the Advanced Computing East China Sub-Center.

% References should be produced using the bibtex program from suitable
% BiBTeX files (here: strings, refs, manuals). The IEEEbib.bst bibliography
% style file from IEEE produces unsorted bibliography list.
% -------------------------------------------------------------------------
\bibliographystyle{IEEEbib}
\bibliography{strings,refs}

\end{document}